\documentclass[conference]{IEEEtran}

\usepackage{setspace}

\usepackage{cite} 

\usepackage{times}
\usepackage{latexsym}

\usepackage{url}

\usepackage{xcolor} 

\definecolor{correctingcolor}{RGB}{0,109,44}
\definecolor{skepticalcolor}{RGB}{116,196,118}
\definecolor{trustingcolor}{RGB}{186,228,179}
\definecolor{questioningcolor}{RGB}{0,90,50}

\definecolor{lineplotscolor}{RGB}{0,109,44}

\definecolor{D}{RGB}{0,0,0}  
\definecolor{V}{RGB}{3,72,23} 
\definecolor{T}{RGB}{3,72,23}  
\colorlet{CB}{blue!75!black} 
\colorlet{CS}{orange!80!black} 
\colorlet{P}{red!75!black} 
 
\colorlet{answer_color}{blue!50!white} 
\colorlet{appreciation_color}{blue!90!white}  
\colorlet{elaborate_color}{blue!75!black} 
\colorlet{humor_color}{blue!40!black}  
\colorlet{question_color}{black} 
\colorlet{no_label_color}{gray}

\colorlet{all_tweets}{black} 
\colorlet{bot}{black!50!white} 
\colorlet{human}{cyan} 

\usepackage{tikz} 
\usepackage{pgfplots}
\pgfplotsset{compat=1.14}

\newcommand*{\ie}{{\em i.e.,~}}
\newcommand*{\eg}{{\em e.g.,~}}
\newcommand*{\etal}{et al.~}
\newcommand*{\ignore}[1]{} 
\newcommand*{\nop}[1]{} 

\usepackage{enumitem}
\setlist[enumerate]{itemsep=0mm}

\makeatletter
\def\@IEEEpubidpullup{2\baselineskip} 
\makeatother

\usepackage{fancyhdr} 
\usepackage{kantlipsum} 
\fancypagestyle{plain}{ 
\fancyhf{} 
\fancyhead[C]{Conference on \LaTeX}  
  
} 
\usepackage{eso-pic} 
\begin{document} 
\AddToShipoutPictureBG*{ 
\AtPageUpperLeft{ 
\setlength\unitlength{1in} 
\hspace*{\dimexpr0.5\paperwidth\relax} 
}
} 
 
\title{How Humans versus Bots React to Deceptive and Trusted News Sources:\,{A\,Case\,Study\,of\,Active\,Users}} 

\author{
\IEEEauthorblockN{Maria Glenski ~~~~~~ Tim Weninger}
\IEEEauthorblockA{Computer Science and Engineering\\University of Notre Dame\\
Notre Dame, Indiana 46556\\
Email: \{mglenski, tweninge\}@nd.edu}\and
\IEEEauthorblockN{Svitlana Volkova}
\IEEEauthorblockA{Data Sciences and Analytics Group \\
  National Security Directorate \\
 Richland, WA 99354 \\
Email: svitlana.volkova@pnnl.gov}
}
 
\maketitle


\begin{abstract}

    Society's reliance on social media as a primary source of news has spawned a renewed focus on the spread of misinformation. In this work, we identify the differences in how social media accounts identified as bots react to news sources of varying credibility, regardless of the veracity of the content those sources have shared. We analyze bot and human responses annotated using a fine-grained model that labels responses as being an answer,  appreciation, agreement, disagreement, an elaboration, humor, or a negative reaction. We present key findings of our analysis into the prevalence of bots, the variety and speed of bot and human reactions, and the disparity in authorship of reaction tweets between these two sub-populations. We observe that bots are responsible for 9-15\% of the reactions to sources of any given type but comprise only 7-10\% of accounts responsible for reaction-tweets; trusted news sources have the highest proportion of humans who reacted; bots respond with significantly shorter delays than humans when posting answer-reactions in response to  sources identified as propaganda. Finally, we report significantly different inequality levels in reaction rates for accounts identified as bots vs not.

\end{abstract}

\section{Introduction}

Misinformation spread in social networks has become a critical focus as users rely on these platforms as a primary source of news~\cite{lazer2018science}. Current studies in this area have focused on rumor and misinformation detection with a primary focus on the network's role in information diffusion models~\cite{qazvinian2011rumor,kwon2013prominent,wu2015false,kwon2017rumor}. 
Other studies compare the behavior of traditional and alternative media~\cite{starbird2017examining}, classify media sources into sub-categories of misinformation~\cite{volkova2017separating}, or attempt to detect rumor-spreading users~\cite{rath2017retweet}. These and other studies have found that the size and shape of (mis)information cascades within a social network depends heavily on the initial reactions of the users. Yet, we still lack an understanding of how users (human and automated alike) react to news sources of varying credibility and how their various response types contribute to the spread of (mis)information. The present work aims to fill this gap by labelling bot and human users' reactions to (mis)information posted by various news sources to measure how bot and human user reactions to deceptive news sources differ from their responses to trusted news sources. 
 
Instead of focusing on user reactions to individual news {\em stories}, the current work compares human-user and bot reactions to news {\em sources} of varying credibility. We focus on how behavior of bot and human users differ in four specific areas: 1) concentration of reactions to news sources of each level of credibility, \ie are bots responsible for a larger proportion of the reactions for one class of news sources over another? \textit{(prevalence of bots)}, 2) the variety of reactions each class of news sources evoke, \textit{(reaction variety)}, 3) the speed with which reactions are posted, \textit{(reaction speed)}, and 4) how equally the volume of reactions are spread across the set of users who reacted \textit{(reaction inequality)}.

\section{Related Work} 

\noindent
\textbf{Prevalence of Bots.}
Previous studies have identified the widespread presence of automated accounts or ``bots'' on social media. A 2014 filing from Twitter acknowledged that 8.5 percent of its active monthly users were automated accounts\footnote{{https://www.sec.gov/Archives/edgar/data/1418091/000156459014003474/ twtr-10q\_20140630.htm?\_ga=1.155500795.1900968760.1407851022}} and subsequent studies found this to be a low estimate of the actual prevalence of bot accounts\cite{varol2017online,chu2010who}. 
Recent work has found that accounts spreading disinformation are significantly more likely to be automated accounts~\cite{shao2017spread}. Other studies highlight evidence of bot participation in political discussion~\cite{howard2016bots,woolley2016automating,metaxas2012social} and astroturf campaigns that present the appearance of widespread support of a candidate, opinion, or topic artificially \cite{ratkiewicz2011detecting}.
A 2018 Pew Research center study found that the majority (66\%) of links tweeted to popular news sites are posted by accounts that are likely to be bots, \ie whose behavior is more similar to bot accounts than to humans~\cite{wojcik2018pew}. We seek to answer whether similar trends hold among reactions to news sources.

\medskip
\noindent
\textbf{Reaction Variety.}
Linguistic markers have been found to be effective for early detection of rumors in social networks. 
For example, Kwon \etal~\cite{kwon2017rumor} demonstrated better detection performance of rumors on Twitter by using user and linguistic features rather than structural or temporal network features. Similarly, Zhao \etal~\cite{zhao2015enquiring} identified clusters of tweets that contain disputed claims by searching for fact-checking language.
Recently, Zhang~\etal~\cite{zhang2017characterizing} classified Reddit comments into eight types including agreement, answer, appreciation, disagreement, elaboration, humor, negative reaction, and question, and analyzed patterns from these discussions arranged by various subreddits. 
Our work goes one step further and employs information credibility classifiers like those mentioned above in order to better understand how (and how fast) human users and bots react to information posted by news sources of varying credibility.

\medskip
\noindent
\textbf{Reaction Speed.} Information diffusion studies have often used epidemiological models, originally formulated to model the spread of disease within a population, in the context of social media \cite{jin2013epidemiological,tambuscio2015fact,wu2016mining}. 
In this context, users are {\it infected} when they spread information to other users. 
A recent study by Vosoughi \etal~\cite{Vosoughi1146} found that news that was fact-checked (post-hoc) and found to be false had spread faster and to more people than news items that were fact-checked and found to be true. In this work, we examine the speed at which users react to content posted by news sources of varying credibility and compare the delays of different types of responses. By contrasting the speed of reactions of different types, from different types of users (bot and human), and in response to sources of varying credibility, we are able to determine whether deceptive or trusted \textit{sources} have slower immediate share-times overall and within each combination of user, reaction, and news source types.

\medskip
\noindent
\textbf{Reaction Inequality.} 
In the context of social media, the 1\% rule and its variants indicate that most users only browse content while a mere 1\% of users contribute new content~\cite{vanmierlo2014rule,hargittai2008participation}. Within the subset of those who actively contribute new content, Kumar and Geethakumari \cite{kumar2014detecting} found a  larger disparity among users who retweeted news from sources that were identified as spreading disinformation. That is, a small number of highly active users were responsible for the vast majority of retweets of disinformation. This study focused only on keywords related to the events in Egypt and Syria in 2013. To answer this research question more generally, the present work quantifies and compares the disparity in sharing behavior of users who frequently reacted to news sources across the various categories of sources, in particular the disparity within each of the reaction types. Specifically, for each type of reaction and each type of news source, we examine whether reactions from bots and human users who frequently reacted are equally distributed across the population of users or if there are a small group of vocal users responsible for the majority of the reaction-tweets.

\section{Data Collection and Annotation} 
 Deceptive news sources that primarily share clickbait, conspiracy theories, or propaganda were previously collected by Volkova et al.~\cite{volkova2017separating} from several public resources that annotate suspicious news accounts.\footnote{Deceptive news lists include \url{http://www.fakenewswatch.com/},  \url{http://www.propornot.com/p/the-list.html}.} The authors also compiled a set of trusted news sources that tweet in English with Twitter-verified accounts which were manually labeled. We collected a set of news sources from https://euvsdisinfo.eu/ that were identified as a source of disinformation by the European Union's East Strategic Communications Task Force. As of November 2016, EUvsDisinfo reports include almost 1,992 confirmed disinformation campaigns found in news reports from around Europe and beyond. We limited our set to news sources identified between 2015 and 2016~\cite{volkova2018misleading}.

 In total, we focused on 282 news sources which were identified as sources who spread:
 \begin{itemize}[noitemsep] 
	\item \textbf{trusted news (T)}: factual information with no intent to deceive the audience;
	\item \textbf{clickbait (CB)}: attention-grabbing, misleading, or vague headlines to attract an audience; 
	\item \textbf{conspiracy theories (CS)}: uncorroborated or unreliable information to explain events or circumstances;
	\item \textbf{propaganda (P)}: intentionally misleading information to advance a social or political agenda; or 
	\item \textbf{disinformation (D)}: fabricated and factually incorrect information meant to intentionally deceive the audience. 
 \end{itemize}
\smallskip

 We collected tweets posted between January 2016 and January 2017 that explicitly @mentioned or directly retweeted content from one of our 282 sources via the public Twitter API and assigned a label to each tweet based on the class of the source @mentioned or retweeted. Then, we focused on the subset of 4,613,517 tweets \nop{posted by 1,021,080 distinct users }identified as English-content in the Twitter metadata. We further focused on users who frequently interacted (at least five times) with the news sources we considered, using tweets posted in any language, which resulted in 431,771 English-tweets for 255 news sources from 184,248 distinct, frequently interacting users. 
 We then classified each of the reaction-tweets as an agreement, answer, appreciation, disagreement, elaboration, humor, negative reaction, question, or other. To do so, we used linguistically-infused neural network models~\cite{glenski2018identifying} trained on a manually annotated reaction dataset from Zhang \etal\cite{zhang2017characterizing}.

  Finally, we gathered botometer scores~\cite{davis2016botornot} for each user who posted a reaction-tweet and partitioned the data into bot reactions and human-user reactions using a bot-score threshold of 0.5. That is, human-user reactions were posted by users with a bot score under the threshold of 0.5 and the bot reactions dataset comprises tweets posted by users with bot scores at or above the threshold. A summary of the dataset across source types is presented in Table~\ref{tab:data_summary}.

  \begin{table}[ht]
 	\centering
 	\small
	 \caption{Summary of English Reactions from users who reacted frequently ($\ge 5$ reactions between Jan 2016 and Jan 2017).} 

 	\begin{tabular}{l@{\hskip0pt}r@{\hskip6pt}rr@{\hskip6pt}r}  
		\hline
    	{} &  \multicolumn{2}{c}{Sources} &  \multicolumn{2}{c}{~Reactions} \\ 
    	Source-Type &  \# Accounts &  \# Tweets  &  \# Users &  \# Tweets\\
        \hline
        Trusted        &        173 &          1,633,996 &               173,098 &            2,875,120 \\
        Clickbait      &         10 &            13,764 &                 8,088 &              22,352 \\
        Conspiracy     &         13 &            31,584 &                14,047 &              80,025 \\
        Propaganda     &         25 &            81,305 &                51,160 &             295,070 \\
        Disinformation &         34 &            68,319 &                26,131 &             164,040 \\

		\hline 
 	\end{tabular}
 	 
 	\label{tab:data_summary}
 \end{table}

\section{Methodology}
In this section, we describe the methodology we used to examine the behavior of bot and human user accounts across varying reactions and in reaction to news sources of varying levels of credibility. As previously discussed, we focus on four specific types of behavior: prevalence of bots, reaction variety, reaction speed, and the inequality of reaction volume.

First, we examine the prevalence of bots, \ie the relative presence of bots in reactions to news sources of each type. We consider the following two distributions: 1) bot scores of users who reacted to news sources of a given type and 2) bot scores associated with reaction-tweets (the bot scores of users who posted the reaction). The distribution of reaction-users focuses on the distribution of bot scores over the set of unique users who reacted, each user is represented once and only once. On the other hand, users may be represented multiple times in the distribution of bot scores associated with reaction-tweets, if they reacted to a news source of a given class multiple times. With these two distributions of bot scores, we are able to examine the prevalence of bots within the population of reacting users and within the population of reactions broadcast. 

As a result of our bot classification methodology, we are able to examine user types using coarse and fine-grained classifications.
We first examine the distributions of bots and humans users at a coarse granularity with a binary classification of users as either a bot or human user account. Then, we consider a fine-grained distinction using the bot scores of users and compare the distributions of bot scores for users who react and of bot scores associated with reaction-tweets (\ie the bot score of the user who posted). Mann Whitney U (MWU) tests that compare distributions across types of sources and types of users are used to identify statistically significant differences in these fine-grained distributions.

The next characteristic that we evaluate is the variety of reactions each class of news source elicits from bots and from human users. We compare distributions across reaction types overall and separated them into each category of user. Comparisons of reaction variety within each user type allows us to identify certain reactions, classes of news sources, or reactions to a class of news source that have higher concentrations of bot (or human) reactions. Then we consider the tendency of each user type by comparing the frequencies of each reaction type across all classes of sources between bot and human users. 

Next we examine the speed of reactions. To answer whether how quickly bots or human users react differs or whether users react to content from trusted sources faster than from deceptive sources, we look at reaction delays for each user type, reaction type, and response to each class of news sources. We define the reaction delay as the time elapsed between the source tweet and when the reaction occurred. We compare the cumulative distribution functions (CDFs) of each user type within and across each type of source to analyze the delay patterns.

	\begin{figure}[t] 
		\centering
		\small
        \usepgfplotslibrary{fillbetween}

        \begin{tikzpicture}
        \begin{axis}[ 
        axis lines = left,
        height=2.25in, width=2.25in,
        xlabel style={yshift=0.05in,align=center}, 
        ylabel style={yshift=-0.1in,align=center}, 
        xlabel={cumulative \% users who reacted\\ \textcolor{gray!50!black}{(cumulative \% of population)}},
        ylabel={\% reaction-tweets\\ \textcolor{gray!50!black}{(\% population's income)}},
        xmin=0-.1,xmax=1.1,ymin=0-.1,ymax=1.1,
        xtick = {0,.25,.5,.75,1},xticklabels ={0,25,50,75,100}, 
        ytick = {0,.25,.5,.75,1},yticklabels ={0,25,50,75,100}, 
        xtick align=center, 
        ytick align=center, 
        title ={Reaction Inequality\\ \textcolor{gray!50!black}{(Income Inequality)}}, title style = {yshift= -.1in, xshift=-.1in,align=center}, 
        ]
         
        \addplot[black,no marks,thick, name path=perfectEquality] coordinates {
        	(0,0) (1,1)
        };

        \addplot[black,no marks,very thick,densely dotted] coordinates {
        	(0.001,0.001) 
        	(1,0.001) (1,1)
        };

        \draw[black!50] (0.65,0.45) node[color=black] {\normalsize $a_1$};
        \draw[black!50] (0.85,0.10) node[anchor=south,color=black] {\normalsize$a_2$};

        \addplot[no marks,black, dashed, thick, name path=lorenz] coordinates {  (0,0) (0.00478687029861,0.00184940554822) (0.00980168680191,0.0037868780273) (0.0148165033052,0.00572435050638) (0.0198313198085,0.00766182298547) (0.0248461363118,0.00959929546455) (0.0298609528151,0.0115367679436) (0.0348757693184,0.0134742404227) (0.0398905858217,0.0154117129018) (0.0449054023251,0.0173491853809) (0.0499202188284,0.01928665786) (0.0549350353317,0.0212241303391) (0.059949851835,0.0231616028181) (0.0649646683383,0.0250990752972) (0.0699794848416,0.0270365477763) (0.0749943013449,0.0289740202554) (0.0797811716435,0.0308234258036) (0.0847959881468,0.0327608982827) (0.0898108046501,0.0346983707618) (0.0948256211534,0.0366358432409) (0.0998404376567,0.0385733157199) (0.10485525416,0.040510788199) (0.109870070663,0.0424482606781) (0.114884887167,0.0443857331572) (0.11989970367,0.0463232056363) (0.124914520173,0.0482606781154) (0.129929336677,0.0501981505945) (0.13494415318,0.0521356230735) (0.139958969683,0.0540730955526) (0.144973786186,0.0560105680317) (0.14998860269,0.0579480405108) (0.154775472988,0.059797446059) (0.159790289492,0.0617349185381) (0.164805105995,0.0636723910172) (0.169819922498,0.0656098634963) (0.174834739002,0.0675473359753) (0.179849555505,0.0694848084544) (0.184864372008,0.0714222809335) (0.189879188512,0.0733597534126) (0.194894005015,0.0752972258917) (0.199908821518,0.0772346983708) (0.204923638021,0.0791721708498) (0.209938454525,0.0811096433289) (0.214953271028,0.083047115808) (0.219968087531,0.0849845882871) (0.224982904035,0.0869220607662) (0.229997720538,0.0888595332453) (0.234784590837,0.0907089387935) (0.23979940734,0.0926464112726) (0.244814223843,0.0945838837517) (0.249829040346,0.0965213562307) (0.25484385685,0.0984588287098) (0.259858673353,0.100396301189) (0.264873489856,0.102333773668) (0.26988830636,0.104271246147) (0.274903122863,0.106208718626) (0.279917939366,0.108146191105) (0.28493275587,0.110083663584) (0.289947572373,0.112021136063) (0.294962388876,0.113958608542) (0.29997720538,0.115896081022) (0.304992021883,0.117833553501) (0.309778892181,0.119682959049) (0.314793708685,0.121620431528) (0.319808525188,0.123557904007) (0.324823341691,0.125495376486) (0.329838158195,0.127432848965) (0.334852974698,0.129370321444) (0.339867791201,0.131307793923) (0.344882607705,0.133245266402) (0.349897424208,0.135182738882) (0.354912240711,0.137120211361) (0.359927057214,0.13905768384) (0.364941873718,0.140995156319) (0.369956690221,0.142932628798) (0.374971506724,0.144870101277) (0.379986323228,0.146807573756) (0.384773193526,0.148656979304) (0.38978801003,0.150594451783) (0.394802826533,0.152531924262) (0.399817643036,0.154469396742) (0.40483245954,0.156406869221) (0.409847276043,0.1583443417) (0.414862092546,0.160281814179) (0.419876909049,0.162219286658) (0.424891725553,0.164156759137) (0.429906542056,0.166094231616) (0.434921358559,0.168031704095) (0.439936175063,0.169969176574) (0.444950991566,0.171906649053) (0.449965808069,0.173844121532) (0.454980624573,0.175781594011) (0.459995441076,0.177719066491) (0.464782311375,0.179568472039) (0.469797127878,0.181505944518) (0.474811944381,0.183443416997) (0.479826760884,0.185380889476) (0.484841577388,0.187318361955) (0.489856393891,0.189255834434) (0.494871210394,0.191193306913) (0.499886026898,0.193130779392) (0.504900843401,0.195068251871) (0.509915659904,0.197005724351) (0.514930476408,0.19894319683) (0.519945292911,0.200880669309) (0.524960109414,0.202818141788) (0.529974925917,0.204755614267) (0.534989742421,0.206693086746) (0.539776612719,0.208542492294) (0.544791429223,0.210479964773) (0.549806245726,0.212417437252) (0.554821062229,0.214354909731) (0.559835878733,0.21629238221) (0.564850695236,0.21822985469) (0.569865511739,0.220167327169) (0.574880328243,0.222104799648) (0.579895144746,0.224042272127) (0.584909961249,0.225979744606) (0.589924777752,0.227917217085) (0.594939594256,0.229854689564) (0.599954410759,0.231792162043) (0.604969227262,0.233729634522) (0.609984043766,0.235667107001) (0.614998860269,0.23760457948) (0.619785730568,0.239453985029) (0.624800547071,0.241391457508) (0.629815363574,0.243328929987) (0.634830180078,0.245266402466) (0.639844996581,0.247203874945) (0.644859813084,0.249141347424) (0.649874629587,0.251078819903) (0.654889446091,0.253016292382) (0.659904262594,0.254953764861) (0.664919079097,0.258828709819) (0.669933895601,0.262703654778) (0.674948712104,0.266578599736) (0.679963528607,0.270453544694) (0.684978345111,0.274328489652) (0.689993161614,0.27820343461) (0.694780031912,0.281902245707) (0.699794848416,0.285777190665) (0.704809664919,0.289652135623) (0.709824481422,0.293527080581) (0.714839297926,0.297402025539) (0.719854114429,0.301276970498) (0.724868930932,0.305151915456) (0.729883747436,0.309026860414) (0.734898563939,0.312901805372) (0.739913380442,0.31677675033) (0.744928196946,0.320651695288) (0.749943013449,0.324526640247) (0.754957829952,0.328401585205) (0.759972646455,0.332276530163) (0.764987462959,0.336151475121) (0.769774333257,0.339850286218) (0.774789149761,0.343725231176) (0.779803966264,0.347600176134) (0.784818782767,0.351475121092) (0.789833599271,0.35535006605) (0.794848415774,0.359225011008) (0.799863232277,0.363099955967) (0.80487804878,0.366974900925) (0.809892865284,0.37155438133) (0.814907681787,0.377366798767) (0.81992249829,0.383179216204) (0.824937314794,0.388991633642) (0.829952131297,0.394804051079) (0.8349669478,0.400616468516) (0.839981764304,0.406428885953) (0.844996580807,0.412241303391) (0.849783451106,0.417789520035) (0.854798267609,0.423601937472) (0.859813084112,0.42941435491) (0.864827900615,0.435226772347) (0.869842717119,0.441039189784) (0.874857533622,0.447203874945) (0.879872350125,0.454953764861) (0.884887166629,0.462703654778) (0.889901983132,0.470453544694) (0.894916799635,0.47820343461) (0.899931616139,0.485953324527) (0.904946432642,0.493703214443) (0.909961249145,0.501453104359) (0.914976065649,0.509202994276) (0.919990882152,0.518185821224) (0.92477775245,0.527432848965) (0.929792568954,0.537120211361) (0.934807385457,0.546807573756) (0.93982220196,0.556847203875) (0.944837018464,0.568472038749) (0.949851834967,0.580096873624) (0.95486665147,0.591985909291) (0.959881467974,0.605548216645) (0.964896284477,0.619638925583) (0.96991110098,0.635755173932) (0.974925917483,0.653192426244) (0.979940733987,0.672831351827) (0.98495555049,0.695728753853) (0.989970366993,0.725319242624) (0.994985183497,0.770849845883) (1,1)
        };

        \addplot[gray!50] fill between[of=perfectEquality and lorenz];
          
        \end{axis}
        \end{tikzpicture}
        
        \begin{tikzpicture} 
        \begin{axis}[%
        hide axis, height =.75in,
        xmin=0,xmax=50,ymin=0,ymax=0.4,
        legend style={at={(0.5,1)},
        	anchor=north,legend columns=-1,column sep=.05cm,draw=none}, 
        ] 
        \addlegendimage{black,solid, thick}
        \addlegendentry{Perfect Equality}; 
        \addlegendimage{black,densely dotted, thick}
        \addlegendentry{Perfect Inequality};  
        \end{axis}
        \end{tikzpicture}
        
        \vspace{-.05in}
        \begin{tikzpicture} 
        \begin{axis}[%
        hide axis, height =.75in,
        xmin=0,xmax=50,ymin=0,ymax=0.4,
        legend style={at={(0.5,1)},
        	anchor=north,legend columns=-1,column sep=.05cm,draw=none,fill=none},   
        ]  
        \addlegendimage{black,dashed, thick}
        \addlegendentry{Example Lorenz Curve}; 
        \end{axis}
        \end{tikzpicture}

		\caption{Lorenz curves and Gini coefficients. 
		As a graphical representation of income inequality within a population, Lorenz curves plot the share of income by the cumulative share of the population. The Gini coefficient is the proportion of the area under the line of perfect equality ($a_1+a_2$) that is captured between the line of perfect equality and the Lorenz curve ($a_1$). We adapt Lorenz curves to measure the inequality in reaction volume by plotting the share of the total reaction volume, \ie the $y$\% of reaction-tweets posted, by the share of the population who reacted, \ie the cumulative $x$\% of users ordered by least to most reaction-tweets posted. }
		\label{fig:lorenz_gini_example} 
	\end{figure}
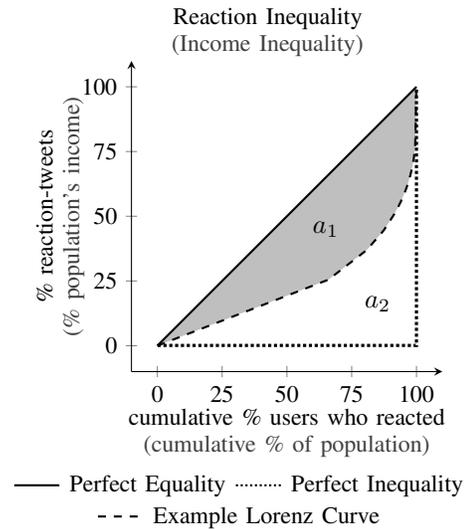

 Finally, we compare the inequality in reactions among bots and human users. That is, how evenly the volume of reaction-tweets is spread across users of each type; Does each user post an equal number of reactions? We do so using two measures that have been commonly used to measure income inequality: Lorenz curves and Gini coefficients. 
 Rather than measure how much of the total population's income each individual is responsible for, we repurpose these metrics to measure how much of the total reaction-tweet volume each user is responsible for. This allows us to compare \textit{reaction inequality} across source types the way that economists compare income inequality across countries or regions. 
	
	Lorenz curves have traditionally been used to illustrate the distribution of income or wealth graphically~\cite{kakwani1973estimation}. In those domains, the curves plot the cumulative percentage of wealth or income compared against the cumulative (in increasing shares) percentage of a corresponding population. The degree to which a Lorenz curve deviates from the straight diagonal line ($y=x$) representative of perfect equality represents the inequality present in the distribution. In our case, the Lorenz curve is adapted to illustrate the cumulative percentage of propagation (tweets shared) as a function of the cumulative percentage of users posting, as shown in Figure~\ref{fig:lorenz_gini_example}.

	\begin{equation}
	\hat{G} =  1 - \sum_{k=1}^{n} (X_k - X_{k-1})(Y_k + Y_{k-1})
	\label{eq:gini}
	\end{equation}
	
	The Gini coefficient is defined as the proportion of the area under the line of perfect equality that is captured above the Lorenz curve, \ie $\frac{a_1}{a_1+a_2}$ in Figure~\ref{fig:lorenz_gini_example}. 
	The Gini coefficients reported in subsequent sections are calculated using the formula in Eq.~\ref{eq:gini}, which is an approximation of the points of the Lorenz curves observed in the collected data. Using income as an example, Gini coefficients can grow larger than 1 but only if individuals within the population can be responsible for negative shares, that is, if individuals can have negative incomes. In our data, users must be responsible for at least 1 reaction-tweet in order to be considered part of the dataset, so Gini coefficients in our analysis have an upper-bound of 1.

 \section{Analysis} 
 Here we present the key results of our analysis of the behavior of bots and human users in reaction to news sources of varying credibility: the prevalence of reactions from bots and the variety, speed, and the inequality in volume of reaction tweets evoked by each class of news source.

 \subsection{Prevalence of Bots}
  
In this subsection, we consider the prevalence of bot users among the audience and reactions broadcast to the community. The distributions of users across bot, human, and unknown (accounts for which we could not collect bot scores) within each class of news source are presented in Table~\ref{tab:bot_prevalence_tab}.

  \begin{table}[t]
 	\centering
 	\small
	 \caption{(Prevalence of Bots) Distributions across bot accounts (bot score $\ge 0.5$), human accounts (bot score $<0.5$), and unknown accounts (for which we could not collect a bot score) within the set of users who reacted (U) and the set of reaction tweets (T) for each class of news source. Highest proportions of each user type are highlighted in bold and lowest proportions are in italics. }
 	 
 	\begin{tabular}{lr@{\hskip6pt}rr@{\hskip6pt}rr@{\hskip6pt}r}  
		\hline 
    	 &	\multicolumn{2}{c}{\% Bot}	&	\multicolumn{2}{c}{\% Human}	&	\multicolumn{2}{c}{\% Unknown}	\\
    	Source-Type	&	\multicolumn{1}{c}{\tiny U}	&	\multicolumn{1}{c}{\tiny T}	&	\multicolumn{1}{c}{\tiny U}	&	\multicolumn{1}{c}{\tiny T}	&	\multicolumn{1}{c}{\tiny U}	&	\multicolumn{1}{c}{\tiny T}	\\
        \hline							
        Trusted	        &	7.47	        &	12.57         &	\textbf{77.32}	&	74.46         &	\textit{15.22}	&	13.03\\
        Clickbait	    &	\textbf{10.17}	&	\textbf{15.06}&	74.10	        &	72.62         &	15.73	                    &	\textit{12.35}\\
        Conspiracy	    &	7.90	        &	 \textit{8.90}&	\textit{72.79}	&	\textbf{76.50}&	\textbf{19.31}	&	14.62\\
        Propaganda	    &	\textit{6.80}	&	11.54         &	75.00	        &	70.56         &	18.20	                    &	\textbf{17.94}\\
        Disinformation	&	9.64	        &	13.29         &	73.11	        &	\textit{70.18}&	17.25	                    &	16.65\\
		\hline  					
		
 	\end{tabular}
 	 
 	\label{tab:bot_prevalence_tab}
 \end{table}
   
\begin{figure}[t]
	\centering 
	\small    
	\small
    \begin{tikzpicture}
    \begin{axis}[
    	ybar,
        height=1.2in, width=1.75in,
    	bar width = 2pt, 
    	ylabel = {\textbf{Trusted}\\\\\% bot scores},
    	ylabel style={align=center},
    	ymin=0,ymax=0.3,ytick ={0,0.1,0.2,0.3}, yticklabels ={0,10,20,30}, 
    	xtick pos = left, 
    	xmin=0.025,xmax=1.025,
    	xtick={0.025, 0.125, 0.225, 0.325, 0.425, 0.525, 0.625, 0.725, 0.825, 0.925, 1.025},
    	xticklabels = {0, ,0.2, ,0.4, ,0.6, ,0.8, ,1}, 
    	tick label style={font=\tiny},
    	title={\textbf{Users}},
    	]
    	
    	\addplot[T, fill=T!75!white] coordinates{
    	(0.05,0.000198067937302) (0.1,0.00325911787743) (0.15,0.0163135955633) (0.2,0.0484906322869) (0.25,0.1058583094) (0.3,0.170041324174) (0.35,0.202713530741) (0.4,0.142824988971) (0.45,0.130796863324) (0.5,0.0794792613867) (0.55,0.0494989781495) (0.6,0.0302143635267) (0.65,0.0116139835964) (0.7,0.00490668299227) (0.75,0.00235880907151) (0.8,0.000972333510394) (0.85,0.000216074113421) (0.9,0.000180061761184) (0.95,6.30216164144e-05)

    	}; 
    \end{axis}
    \end{tikzpicture}
    \hspace{-0.1in} 
    \begin{tikzpicture}
    \begin{axis}[
    	ybar,  
        height=1.2in, width=1.75in,
    	bar width = 2pt, 
    	xmin=-.15,xmax=1, 
    	ymin=0,ymax=0.3,ytick ={0,0.1,0.2,0.3}, yticklabels ={0,10,20,30}, 
    	xtick pos = left, 
    	xmin=0.025,xmax=1.025,
    	xtick={-0.1,0.025, 0.125, 0.225, 0.325, 0.425, 0.525, 0.625, 0.725, 0.825, 0.925, 1.025},
    	xticklabels = {?,0, ,0.2, ,0.4, ,0.6, ,0.8, ,1},
    	title={\textbf{Reactions}},
    	tick label style={font=\tiny},
    	] 
    	\addplot[T, fill=T!75!white] coordinates{
    		
    	(0.05,0.00010929959379) (0.1,0.00145812864682) (0.15,0.00835120962221) (0.2,0.0310014485199) (0.25,0.0717774035701) (0.3,0.12378118942) (0.35,0.173271084612) (0.4,0.128089515166) (0.45,0.186523960633) (0.5,0.105869268077) (0.55,0.0736150780591) (0.6,0.0568898380204) (0.65,0.0202876861396) (0.7,0.0105227883648) (0.75,0.00233732977489) (0.8,0.00565715479945) (0.85,0.000230610131952) (0.9,0.000117707254851) (0.95,0.00010929959379)
    
    	}; 
    \end{axis}
    \end{tikzpicture} 
    
    \vspace{0.05in}
    
    \begin{tikzpicture}
    \begin{axis}[
    	ybar,
        height=1.2in, width=1.75in,
    	bar width = 2pt, 
    	xmin=-.15,xmax=1,
    	ymin=0,ymax=0.3,ytick ={0,0.1,0.2,0.3}, yticklabels ={0,10,20,30},
    	xtick pos = left, 
    	xmin=0.025,xmax=1.025,
    	xtick={-0.1,0.025, 0.125, 0.225, 0.325, 0.425, 0.525, 0.625, 0.725, 0.825, 0.925, 1.025},
    	xticklabels = {?,0, ,0.2, ,0.4, ,0.6, ,0.8, ,1},
    	ylabel = {\% unique users},
    	ylabel style ={align=center},
    	ylabel = {\textbf{Clickbait}\\\\\% bot scores},
    	tick label style={font=\tiny},
    	]
    	\addplot[CB, fill=CB!75!white] coordinates{
        	
    	(0.1,0.00101996211569) (0.15,0.00684831706251) (0.2,0.0212734955559) (0.25,0.0606148914469) (0.3,0.141046189713) (0.35,0.200349701297) (0.4,0.162173976395) (0.45,0.162902520764) (0.5,0.104036135801) (0.55,0.0700859682355) (0.6,0.0492495993006) (0.65,0.0126766720093) (0.7,0.00568264607315) (0.75,0.00131137986303) (0.8,0.000582835494682) (0.95,0.00014570887367)

    	}; 
    \end{axis}
    \end{tikzpicture}
    \hspace{-0.1in} 
    \begin{tikzpicture}
    \begin{axis}[
    	ybar,
        height=1.2in, width=1.75in,
    	bar width = 2pt, 
    	xmin=-.15,xmax=1, 
    	ymin=0,ymax=0.3,ytick ={0,0.1,0.2,0.3}, yticklabels ={0,10,20,30},
    	xtick pos = left, 
    	xmin=0.025,xmax=1.025,
    	xtick={-0.1,0.025, 0.125, 0.225, 0.325, 0.425, 0.525, 0.625, 0.725, 0.825, 0.925, 1.025},
    	xticklabels = {?,0, ,0.2, ,0.4, ,0.6, ,0.8, ,1}, 
    	tick label style={font=\tiny},
    	]
    	\addplot[CB, fill=CB!75!white] coordinates{
    		
    	(0.1,0.000700350175088) (0.15,0.00435217608804) (0.2,0.0182591295648) (0.25,0.0494247123562) (0.3,0.111655827914) (0.35,0.178689344672) (0.4,0.172386193097) (0.45,0.163581790895) (0.5,0.108254127064) (0.55,0.0914457228614) (0.6,0.0745872936468) (0.65,0.0139569784892) (0.7,0.00940470235118) (0.75,0.00245122561281) (0.8,0.000600300150075) (0.95,0.000250125062531)
    
    	}; 
    \end{axis}
    \end{tikzpicture} 
    
    \vspace{0.05in}
    
    \begin{tikzpicture}
    \begin{axis}[
    	ybar,
        height=1.2in, width=1.75in,
    	bar width = 2pt,  
    	ymin=0,ymax=0.3,ytick ={0,0.1,0.2,0.3}, yticklabels ={0,10,20,30},
    	xtick pos = left, 
    	xmin=0.025,xmax=1.025,
    	xtick={-0.1,0.025, 0.125, 0.225, 0.325, 0.425, 0.525, 0.625, 0.725, 0.825, 0.925, 1.025},
    	xticklabels = {?,0, ,0.2, ,0.4, ,0.6, ,0.8, ,1},
    	ylabel = {\% unique users},
    	ylabel style ={align=center},
    	ylabel = {\textbf{Conspiracy}\\\\\% bot scores},
    	tick label style={font=\tiny},
    	]
    	\addplot[CS, fill=CS!75!white] coordinates{

    	(0.05,0.000173913043478) (0.1,0.00113043478261) (0.15,0.0084347826087) (0.2,0.0331304347826) (0.25,0.085652173913) (0.3,0.157739130435) (0.35,0.210086956522) (0.4,0.15347826087) (0.45,0.147652173913) (0.5,0.0899130434783) (0.55,0.0584347826087) (0.6,0.0325217391304) (0.65,0.0113913043478) (0.7,0.00634782608696) (0.75,0.0024347826087) (0.8,0.000869565217391) (0.85,0.000173913043478) (0.9,0.000173913043478) (0.95,0.000260869565217)

    	}; 
    \end{axis}
    \end{tikzpicture}
    \hspace{-0.1in} 
    \begin{tikzpicture}
    \begin{axis}[
    	ybar,
        height=1.2in, width=1.75in,
    	bar width = 2pt, 
    	xmin=-.15,xmax=1, 
    	ymin=0,ymax=0.3,ytick ={0,0.1,0.2,0.3}, yticklabels ={0,10,20,30},
    	xtick pos = left, 
    	xmin=0.025,xmax=1.025,
    	xtick={-0.1,0.025, 0.125, 0.225, 0.325, 0.425, 0.525, 0.625, 0.725, 0.825, 0.925, 1.025},
    	xticklabels = {?,0, ,0.2, ,0.4, ,0.6, ,0.8, ,1}, 
    	tick label style={font=\tiny},
    	]
    	\addplot[CS, fill=CS!75!white] coordinates{
    		
    	(0.05,4.21828203434e-05) (0.1,0.00108269238881) (0.15,0.00805691868558) (0.2,0.029232694498) (0.25,0.0883027039188) (0.3,0.235028614013) (0.35,0.186026237714) (0.4,0.155809277408) (0.45,0.128910698969) (0.5,0.0562296995177) (0.55,0.0483133902333) (0.6,0.0278266004865) (0.65,0.0192634879568) (0.7,0.012401749181) (0.75,0.00185604409511) (0.8,0.00134985025099) (0.85,5.62437604578e-05) (0.9,0.00012654846103) (0.95,8.43656406867e-05)

    	}; 
    \end{axis}
    \end{tikzpicture}
    
    \vspace{0.05in}
    
    \begin{tikzpicture}
    \begin{axis}[
    	ybar,
        height=1.2in, width=1.75in,
    	bar width = 2pt,  
    	ymin=0,ymax=0.3,ytick ={0,0.1,0.2,0.3}, yticklabels ={0,10,20,30},
    	xtick pos = left, 
    	xmin=0.025,xmax=1.025,
    	xtick={-0.1,0.025, 0.125, 0.225, 0.325, 0.425, 0.525, 0.625, 0.725, 0.825, 0.925, 1.025},
    	xticklabels = {?,0, ,0.2, ,0.4, ,0.6, ,0.8, ,1},
    	ylabel = {\% unique users},
    	ylabel style ={align=center},
    	ylabel = {\textbf{Propaganda}\\\\\% bot scores},
    	tick label style={font=\tiny},
    	]
    	\addplot[P, fill=P!75!white] coordinates{

    	(0.05,0.000140947637953) (0.1,0.00216119711527) (0.15,0.0111113721253) (0.2,0.0366228945947) (0.25,0.0922502290399) (0.3,0.160656815993) (0.35,0.213888040593) (0.4,0.155253823205) (0.45,0.144330381263) (0.5,0.0877399046254) (0.55,0.0509995536658) (0.6,0.0299983556109) (0.65,0.0091850877399) (0.7,0.00328877821889) (0.75,0.00129202001456) (0.8,0.000610773097794) (0.85,0.000234912729921) (0.9,0.000140947637953) (0.95,9.39650919683e-05)

    	}; 
    \end{axis}
    \end{tikzpicture}
    \hspace{-0.1in} 
    \begin{tikzpicture}
    \begin{axis}[
    	ybar,
        height=1.2in, width=1.75in,
    	bar width = 2pt, 
    	xmin=-.15,xmax=1, 
    	ymin=0,ymax=0.3,ytick ={0,0.1,0.2,0.3}, yticklabels ={0,10,20,30},
    	xtick pos = left, 
    	xmin=0.025,xmax=1.025,
    	xtick={-0.1,0.025, 0.125, 0.225, 0.325, 0.425, 0.525, 0.625, 0.725, 0.825, 0.925, 1.025},
    	xticklabels = {?,0, ,0.2, ,0.4, ,0.6, ,0.8, ,1}, 
    	tick label style={font=\tiny},
    	]
    	\addplot[P, fill=P!75!white] coordinates{
    			
    	(0.05,7.28697736179e-05) (0.1,0.000817760792823) (0.15,0.008031868381) (0.2,0.0263262298798) (0.25,0.0691129319558) (0.3,0.125769180944) (0.35,0.198833273958) (0.4,0.152706707258) (0.45,0.158507950902) (0.5,0.104470965444) (0.55,0.0636679405383) (0.6,0.0781973637335) (0.65,0.00857029504162) (0.7,0.00246542734074) (0.75,0.00169219807624) (0.8,0.000445315283221) (0.85,0.000109304660427) (0.9,5.66764905917e-05) (0.95,0.000145739547236)

    	}; 
    \end{axis}
    \end{tikzpicture} 
    
    \vspace{-0.05in}
    
    \begin{tikzpicture}
    \begin{axis}[
    	ybar,
        height=1.2in, width=1.75in,
    	bar width = 2pt,  
    	ymin=0,ymax=0.3,ytick ={0,0.1,0.2,0.3}, yticklabels ={0,10,20,30},
    	xtick pos = left, 
    	xmin=0.025,xmax=1.025, xlabel={Bot Score Bin},
    	xtick={-0.1,0.025, 0.125, 0.225, 0.325, 0.425, 0.525, 0.625, 0.725, 0.825, 0.925, 1.025},
    	xticklabels = {?,0, ,0.2, ,0.4, ,0.6, ,0.8, ,1},
    	ylabel = {\% unique users},
    	ylabel style ={align=center},
    	ylabel = {\textbf{Disinformation}\\\\\% bot scores},
    	tick label style={font=\tiny},
    	]
    	\addplot[D, fill=D!75!white] coordinates{
    			
    	(0.05,0.000137734722924) (0.1,0.00160690510078) (0.15,0.0102382810707) (0.2,0.0321381020155) (0.25,0.0846609430237) (0.3,0.147513888251) (0.35,0.201827280657) (0.4,0.15325283504) (0.45,0.144345989624) (0.5,0.0939350810339) (0.55,0.0583995225196) (0.6,0.0401267159451) (0.65,0.0172168403655) (0.7,0.00794270235526) (0.75,0.00454524585648) (0.8,0.00128552408062) (0.85,0.000459115743079) (0.9,0.000275469445847) (0.95,9.18231486158e-05)

    	}; 
    \end{axis}
    \end{tikzpicture}
    \hspace{-0.1in}  
    \begin{tikzpicture}
    \begin{axis}[
    	ybar,
        height=1.2in, width=1.75in,
    	bar width = 2pt, 
    	xmin=-.15,xmax=1, xlabel={Bot Score Bin},
    	ymin=0,ymax=0.3,ytick ={0,0.1,0.2,0.3}, yticklabels ={0,10,20,30},
    	xtick pos = left, 
    	xmin=0.025,xmax=1.025,
    	xtick={-0.1,0.025, 0.125, 0.225, 0.325, 0.425, 0.525, 0.625, 0.725, 0.825, 0.925, 1.025},
    	xticklabels = {?,0, ,0.2, ,0.4, ,0.6, ,0.8, ,1}, 
    	tick label style={font=\tiny},
    	]
    	\addplot[D, fill=D!75!white] coordinates{
    	(0.05,7.22381547486e-05) (0.1,0.0010907961367) (0.15,0.00480383729078) (0.2,0.0151555648663) (0.25,0.0619947844052) (0.3,0.106392354314) (0.35,0.176355007188) (0.4,0.168798896201) (0.45,0.162391371875) (0.5,0.112684297592) (0.55,0.102729879868) (0.6,0.0444481366168) (0.65,0.0217292369484) (0.7,0.00705044390346) (0.75,0.0051433566181) (0.8,0.00825682108776) (0.85,0.000534562345139) (0.9,0.000353966958268) (0.95,1.44476309497e-05)
    
    	};  
    \end{axis}
    \end{tikzpicture}
  
    \vspace{-.1in}
	\caption{(Prevalence of Bots) Bot score distributions, using a bin width of 0.05, for users who reacted (left) and reaction-tweets (right). Mann Whitney U comparisons of raw distributions found that the average bot score of a user who posted a reaction-tweet is higher ($p<0.01$) than the average bot score of a user who reacted for all source types except for Conspiracy-sources, where the average bot score of a user who posted a reaction-tweet is lower ($p<0.01$).}
	\label{fig:dist_bot_score_bin}
\end{figure}
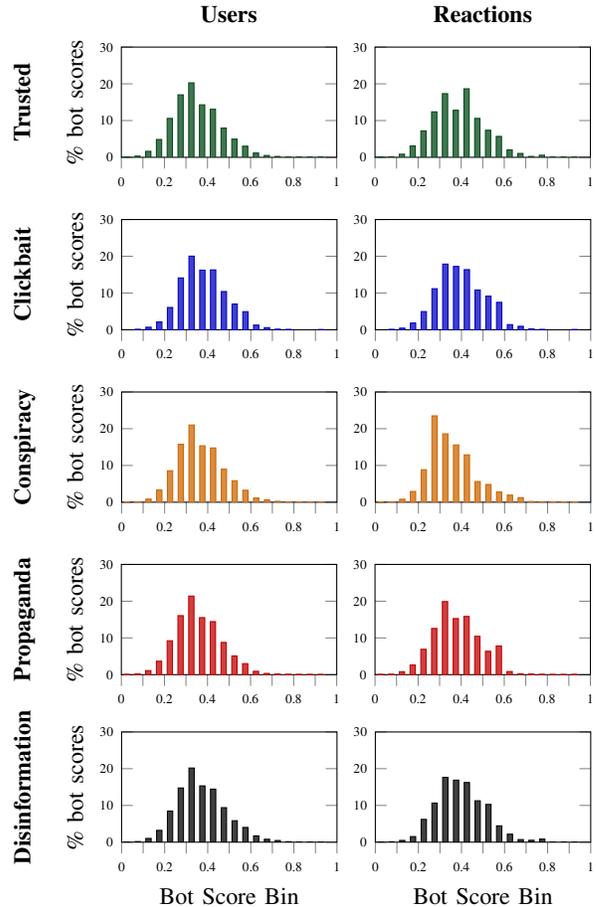

 As shown in Table~\ref{tab:bot_prevalence_tab}, bots are responsible for approximately 9-15\% of the reactions to sources of any given type but only comprise around 7-10\% of users responsible for reaction-tweets. We see that although conspiracy sources have the lowest presence of human users within the population of users who react, they have the highest proportion of reactions authored by human-users. \textit{Trusted news sources have the highest relative presence of human users.} Interestingly, disinformation news sources have only the second highest proportions of bots for users who reacted as well as reaction tweets posted. Instead, clickbait news sources have the highest presence of bots with 10.17\% of users who were responsible for 15.06\% of the reaction-tweets for clickbait sources identified as bots.

 Figure~\ref{fig:dist_bot_score_bin} illustrates the distributions of bot scores of users who reacted (left) and the scores associated with reaction-tweets, \ie the bot score of the user who posted the tweet, (right). When we compare distributions of users' bot scores across classes of news sources, we find statistically significant differences. Mann Whitney U comparisons identified significant ($p<0.01$) differences between distributions for clickbait and trusted or propaganda news sources, where reactions and users who post reactions to clickbait sources have higher bot scores, on average, than trusted or propaganda news sources.
  Although the distributions of bot scores of unique users and scores associated with reaction tweets are not statistically significant, the slight changes in the shape of the distributions, \eg between the two distributions for Conspiracy sources, paired with the discrepancies in Table~\ref{tab:bot_prevalence_tab} hint at the inequality of reaction tweet volume. That is, they indicate that reactions are not evenly spread across users. We investigate this further in our analysis of reaction inequality.

 \subsection{Reaction Variety}

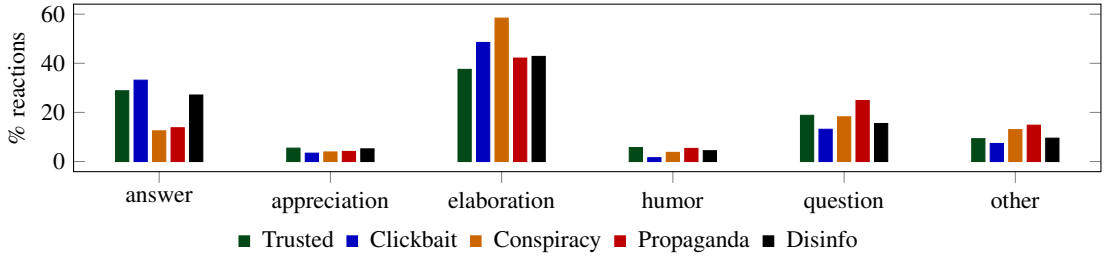
\begin{figure*}[h]
	\centering 
	\small    
	\small
    \begin{tikzpicture}
    \begin{axis}[
    	ybar,  
    	title style={yshift=-.1in},
        height=1.5in,width=6in,
    	bar width = 5pt,
    	ylabel = {\% reactions},
    	ytick = {0,0.2,0.4,0.6}, yticklabels={0,20,40,60},
    	symbolic x coords = {Answer,Appreciation,Elaboration,Humor,Question,Other},
    	xtick={Answer,Appreciation,Elaboration,Humor,Question,Other},
    	xticklabels={answer,appreciation,elaboration,humor,question,other},
    	xtick pos = left,
    	]
    	 
    	\addplot[T,fill=T] coordinates{
            (Answer,	0.288164636	)
            (Appreciation,	0.054556366	)
            (Elaboration,	0.375264733	)
            (Humor,	0.056919776	)
            (Question,	0.188214487	)
            (Other,	0.093092766	)
    	};
    	\addplot[CB,fill=CB] coordinates{ 
    (Answer,	0.331222601	)	
    (Appreciation,	0.034052163	)	
    (Elaboration,	0.484878006	)	
    (Humor,	0.015631227	)	
    (Question,	0.131337732	)	
    (Other,	0.073462339	)	
    
    	};
    	\addplot[CS,fill=CS] coordinates{ 
        (Answer,	0.124920518	)	
        (Appreciation,	0.038945953	)	
        (Elaboration,	0.583981414	)	
        (Humor,	0.037258498	)	
        (Question,	0.182098313	)	
        (Other,	0.129787234	)	
    
    	};
    	\addplot[P,fill=P] coordinates{ 
    			
    (Answer,	0.137570173	)	
    (Appreciation,	0.041100723	)	
    (Elaboration,	0.42125721	)	
    (Humor,	0.053416829	)	
    (Question,	0.248240436	)	
    (Other,	0.148068763	)	
    
    	};
    	\addplot[D,fill=D] coordinates{ 
    			
    (Answer,	0.270801862	)
    (Appreciation,	0.051581889	)
    (Elaboration,	0.428018372	)
    (Humor,	0.043967537	)
    (Question,	0.154849639	)
    (Other,	0.095139658	)
    
    	};
    	
    \end{axis}
    \end{tikzpicture}

    \begin{tikzpicture} 
    \begin{axis}[%
    hide axis,
	height=1in, width=2in,
    xmin=0,xmax=50,ymin=0,ymax=0.4,
    legend style={at={(1.5,0)},draw=none,fill=none, legend cell align=left,legend columns = -1, column sep = 1mm}
    ]
    \addlegendimage{only marks,mark=square*,T, fill=T}
    \addlegendentry{Trusted};
    \addlegendimage{only marks,mark=square*,CB, fill=CB}
    \addlegendentry{Clickbait};
    \addlegendimage{only marks,mark=square*,CS, fill=CS}
    \addlegendentry{Conspiracy}; 
    \addlegendimage{only marks,mark=square*,P, fill=P}
    \addlegendentry{Propaganda};
    \addlegendimage{only marks,mark=square*,D, fill=D}
    \addlegendentry{Disinfo};
    \end{axis}
    \end{tikzpicture}
    
    \vspace{-.1in}
	\caption{(Reaction Variety) Distributions of predicted reaction-types within tweets that directly responded to sources of each source-type. } 
	\label{fig:dist_by_LABEL}
\end{figure*}

  \begin{table*}[t]
 	\centering
 	\small
	 \caption{(Reaction Variety) Proportions of reactions posted by bot, human, or unknown users for each source class and reaction type combination for the most frequent reaction types. Source class(es) with the lowest proportions for each user type are highlighted with bold for each of the reaction types.}
 	 
 	\begin{tabular}{lrrr@{\hskip20pt}rrr@{\hskip20pt}rrr@{\hskip20pt}rrr}
		\hline																									
	&	\multicolumn{3}{c}{Answer}			&	\multicolumn{3}{c}{Elaboration}			&	\multicolumn{3}{c}{Question}			&	\multicolumn{3}{c}{Other}			\\

	&	\multicolumn{1}{c}{\tiny B}	&	\multicolumn{1}{c}{\tiny H}	&	\multicolumn{1}{c}{\tiny U}	&	\multicolumn{1}{c}{\tiny B}	&	\multicolumn{1}{c}{\tiny H}	&	\multicolumn{1}{c}{\tiny U}	&	\multicolumn{1}{c}{\tiny B}	&	\multicolumn{1}{c}{\tiny H}	&	\multicolumn{1}{c}{\tiny U}	&	\multicolumn{1}{c}{\tiny B}	&	\multicolumn{1}{c}{\tiny H}	&	\multicolumn{1}{c}{\tiny U}	\\ 

    \hline																									
    Trusted	&	0.16	&	0.69	&	0.15	&	0.10	&	0.77	&	\textbf{0.13}	&	0.12	&	0.75	&	\textbf{0.13}	&	0.12	&	0.74	&	0.14	\\
    
    Clickbait	&	0.24	&	0.66	&	\textbf{0.11}	&	0.11	&	0.76	&	\textbf{0.13}	&	0.12	&	0.75	&	\textbf{0.13}	&	0.12	&	0.76	&	\textbf{0.12}	\\
    
    Conspiracy	&	\textbf{0.09}	&	0.75	&	0.16	&	\textbf{0.09}	&	0.77	&	0.14	&	\textbf{0.08}	&	0.79	&	\textbf{0.13}	&	\textbf{0.09}	&	0.74	&	0.17	\\
    
    Propaganda	&	0.25	&	\textbf{0.57}	&	0.17	&	\textbf{0.09}	&	0.73	&	0.18	&	0.10	&	\textbf{0.71}	&	0.19	&	\textbf{0.09}	&	0.73	&	0.19	\\
    
    Disinformation	&	0.15	&	0.68	&	0.17	&	0.12	&	\textbf{0.71}	&	0.16	&	0.12	&	\textbf{0.71}	&	0.16	&	0.14	&	\textbf{0.70}	&	0.16	\\
    
    \hline			 
 	\end{tabular}  
 	\label{tab:bot_human_breakdown}
 \end{table*}

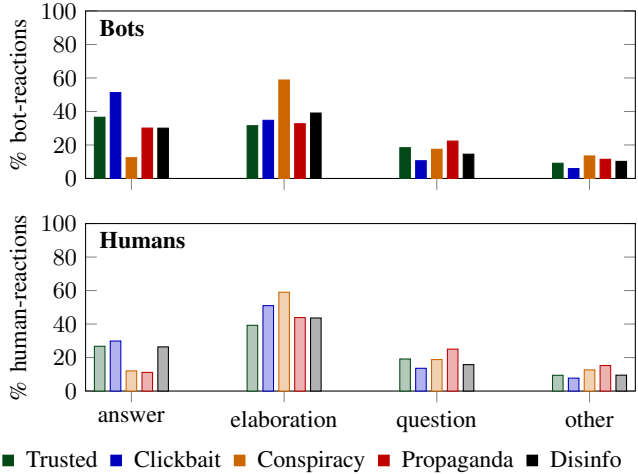
\begin{figure}[h]
	\centering 
	\small    
	\small
\begin{tikzpicture}
\begin{axis}[
	ybar,
	ymin=0,
	ymax=100, 
    height=1.5in, width=3.5in,
	bar width = 4pt, 
	ylabel = {\% bot-reactions\\}, 
	ylabel style = {align=center, yshift = -0.15 in},
	symbolic x coords ={answer,elaboration,question,other}, 
	xtick={answer,elaboration,question,other},
	xticklabels={,,,,,,,,},
	xtick pos = left,  
	]  
	\node[anchor=west] at (rel axis cs:0.01,0.9) { \textbf{Bots} };
	\addplot[T, fill=T] coordinates{
	(answer,36.5701151698) 
    (elaboration,31.5019854019)  
    (question,18.356317862)  
    (other,9.08589128895)  
    };
	\addplot[CB, fill=CB] coordinates{
    (answer,51.2658227848) 	 
    (elaboration,34.7238204833)  
    (question,10.586881473)  
    (other,5.89758342923)
    };
	\addplot[CS, fill=CS] coordinates{
    (answer,12.4150122382)  
    (elaboration,58.7843350558)  
    (question,17.4190916508)  
    (other,13.4484634213)  
    };
	\addplot[P, fill=P] coordinates{
    (answer,30.089768673)  
    (elaboration,32.6706180228)  
    (question,22.3270802164) 	 
    (other,11.4397514098)  
    };
	\addplot[D, fill=D] coordinates{
    (answer,30.0242565807) 	 
    (elaboration,39.0620788788)  
    (question,14.5000449196) 	
    (other,10.2012397808) 	 
	} ;

\end{axis}
\end{tikzpicture}


\begin{tikzpicture}
\begin{axis}[
	ybar,
	ymin=0,
	ymax=100,  
    height=1.5in, width=3.5in,
	bar width = 4pt, 
	ylabel = {\% human-reactions\\}, 
	ylabel style = {align=center, yshift = -0.15 in},
	symbolic x coords ={answer,elaboration,question,other}, 
	xtick={answer,elaboration,question,other}, 
	xtick pos = left, 
	ytick ={0,20,40,60,80,100},
	]  
	\node[anchor=west] at (rel axis cs:0.01,0.9) { \textbf{Humans} };
	\addplot[T, fill=T!30!white] coordinates{
	 (answer,26.7358084129) 	
    (elaboration,39.249658322) 	
    (question,19.1384653457) 	
    (other,9.41807884184) 	
    };
	\addplot[CB, fill=CB!30!white] coordinates{
    (answer,29.8716240765) 	
    (elaboration,50.9870412983) 	
    (question,13.5884703888) 	
    (other,7.71466634371) 	
    };
	\addplot[CS, fill=CS!30!white] coordinates{
	(answer,12.0489296636) 	
    (elaboration,58.9884733004) 	
    (question,18.8002822865) 	
    (other,12.6244805144) 	
    };
	\addplot[P, fill=P!30!white] coordinates{
    
    (answer,11.1383304989) 	
    (elaboration,43.872575577) 	
    (question,25.0335662161) 	
    (other,15.2491908775) 	
    };
	\addplot[D, fill=D!30!white] coordinates{
    
    (answer,26.3706088439) 	
    (elaboration,43.6072685954) 	
    (question,15.7638309044) 	
    (other,9.48773790361) 	
	};

\end{axis}
\end{tikzpicture}

    \begin{tikzpicture} 
    \begin{axis}[%
    hide axis,
	height=1in, width=2in,
    xmin=0,xmax=50,ymin=0,ymax=0.4,
    legend style={at={(1.5,0)},draw=none,fill=none, legend cell align=left,legend columns = -1, column sep = 1mm}
    ]
    \addlegendimage{only marks,mark=square*,T, fill=T}
    \addlegendentry{Trusted};
    \addlegendimage{only marks,mark=square*,CB, fill=CB}
    \addlegendentry{Clickbait};
    \addlegendimage{only marks,mark=square*,CS, fill=CS}
    \addlegendentry{Conspiracy}; 
    \addlegendimage{only marks,mark=square*,P, fill=P}
    \addlegendentry{Propaganda};
    \addlegendimage{only marks,mark=square*,D, fill=D}
    \addlegendentry{Disinfo};
    \end{axis}
    \end{tikzpicture}
    \vspace{-.1in}
	\caption{(Reaction Variety) Frequencies of most common reaction-types within reactions to news sources of each class posted by bot accounts (above) and human user accounts (below), as a percentage of reactions posted by accounts within each population. } 
	\label{fig:reaction_variety_usertype}
\end{figure}

  We plot the distributions of reaction-types for each of the five classes of news sources in Figure~\ref{fig:dist_by_LABEL} and the distribution across bot, human, and unknown users for each source class and reaction type combination for the most frequent reaction types in Table~\ref{tab:bot_human_breakdown}. When we compare the distributions of reaction types, we see that the most common reaction types (\ie present in $\ge 10\%$ of reactions) are answer, elaboration, question, and ``other'' across all classes of media. 
  In Figure~\ref{fig:reaction_variety_usertype} we present the relative frequencies of the most common reactions within the reaction-tweets posted by a given user type in response to news sources of a given class. These plots focus more closely on how reaction frequencies differ within a single user-type population.
  
  When we examine the distributions of each class, we find several key differences in the variety of reactions elicited. Conspiracy news sources have the highest relative rate of elaboration responses, \ie {\it ``On the next day, radiation level has gone up. $[url]$''}, with a more pronounced difference within the bot population. Conspiracy news sources also have the lowest relative rate of answer reactions within the bot population, but not within human users. Clickbait news sources, on the other hand, have the highest relative rate of answer reactions and the lowest rate of question reactions across both populations of user types. 
  
  Conspiracy and propaganda news sources have higher rates of human question-reactions than they do human answer-reactions; human users who react to these types of news sources question content from the source more often than they respond with an answer. While we see a similar trend within human users for conspiracy sources, we see a higher relative rate of answer reactions to propaganda sources when we examine relative rates of bot reactions.

 \subsection{Reaction Speed} 
\begin{figure*}[t]
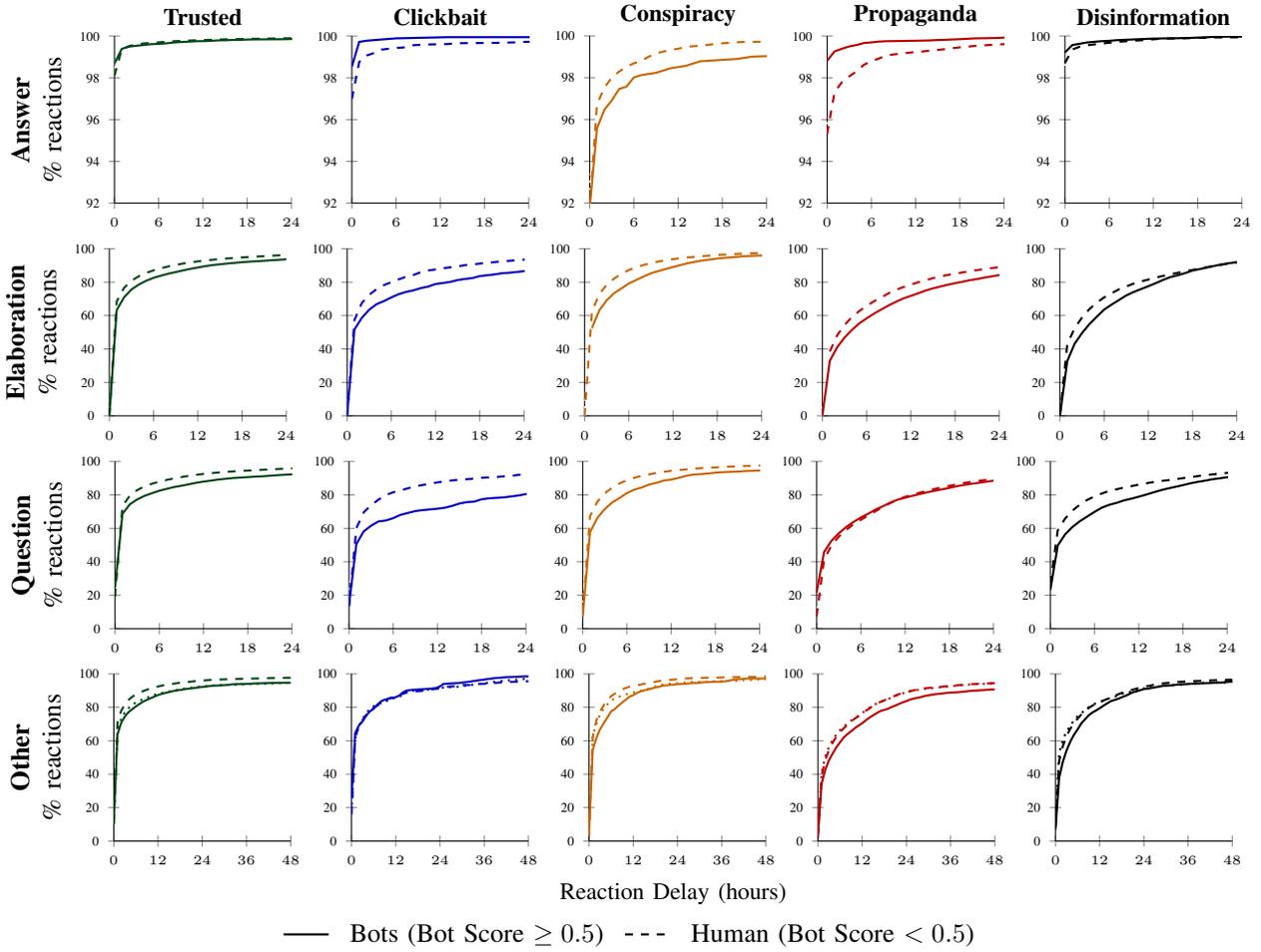

	\centering   


	\vspace{-.1in}
	\caption{(Reaction Speed) Cumulative distribution function (CDF) plots of the volumes of reactions by reaction delays in hours (\ie the delay between when a source posted content and when the reaction tweet was posted) for bots and human user accounts for the most frequently occurring reactions (occurring in at least 10\% of tweets) for each source-type, using a step size of one day. }       
	\label{fig:diffusion_delays_hours} 
\end{figure*}

Next, we study the speed with which bot and human users react to news sources. CDF plots for reaction delays of the most frequently occurring reactions are shown in Figure~\ref{fig:diffusion_delays_hours}. These plots illustrate the percentage of reactions that occur within the first $x$ hours after a source posted the original content users reacted to. As expected, a large proportion of the reaction activity occurs soon after a news source posts across all reaction and source type combinations.

Mann Whitney U tests that compared distributions of reaction delays found that humans elaborate on and question content from clickbait sources faster than bots do ($p<0.01$). This is reflected in Figure~\ref{fig:diffusion_delays_hours} where we see the CDF curve for humans pulls above the curve for bots due to the heavier concentration (at least 80\%) of reactions with very short ($\le 6$ hours) delays, compared to bot users with approximately 60-70\% of reactions that occurred within the first 6 hours. 
We see similar trends for all the other combinations of reaction and source types but a few notable exceptions. \textit{In the case of answer-reactions in response to content from propaganda news sources, bots respond with significantly shorter delays than human users do ($p<0.01$)}. MWU tests comparing bot and human answer-reactions to clickbait and disinformation sources were not found to differ with statistical significance.

\subsection{Reaction Inequality} 
\label{subsection:reaction_inequality}

\begin{figure*}[t]
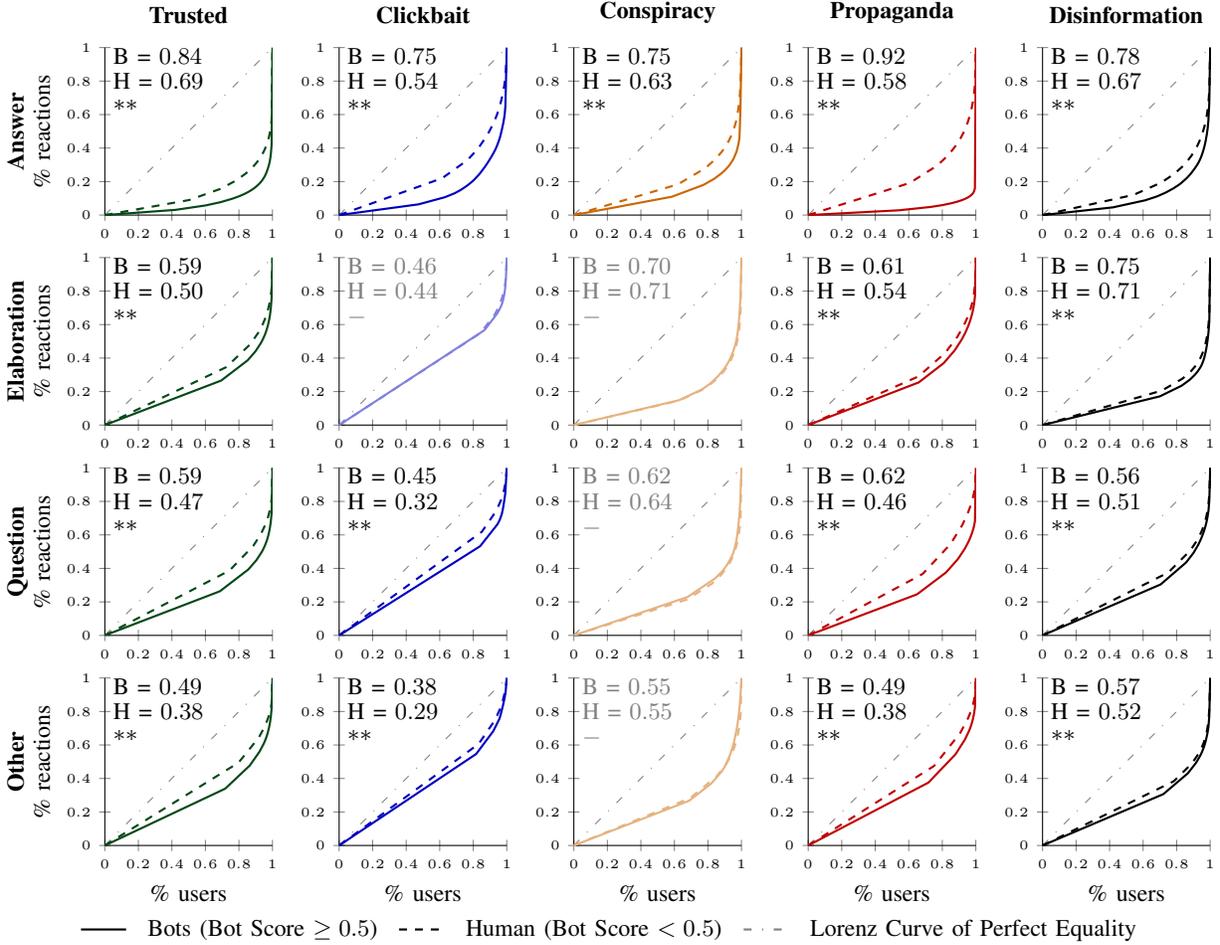

	\centering   
	\small



	\vspace{-.1in}
    
	\caption{(Reaction Inequality) Lorenz curves for each of the frequently occurring reactions (occurring in at least 5\% of tweets) for each source-type. These Lorenz curves plot the share of reactions by the cumulative share of the population (bots, humans, or accounts without bot scores) as a graphical representation of inequality in reaction volume within each population.
	The gray dash-dotted line reflects the Lorenz curve that would result from a population wherein each user was responsible for an equal number of reactions. Gini coefficients for bot (B) and human (H) accounts and statistical significance results of Mann Whitney U (MWU) comparisons of Lorenz curves are listed in the top left corner of each subplot. ** if  $p<0.01$, * if $p<0.05$, and $-$ if $p\ge0.05$. Lorenz curves and Gini coefficients are presented faded where there are no significant differences between bot an human users. }      
	\label{fig:lorenz} 
\end{figure*}

  \begin{table*}[!h] 
 	\centering
 	\small
	 \caption{(Reaction Inequality) The difference ($\Delta$) if statistically significant (MWU $p<0.01$) between Gini coefficients for bot (B) and human (H) user accounts and the relative increase (\%$\Delta$) from the human user to bot Gini coefficient, \ie (B-H)/H. A dash ($-$) is shown if no significant difference was found ($p\ge 0.05$). Highest relative increases are highlighted in bold within source types and italicized within reaction types.}

 	\begin{tabular}{lrrrrrrrrrr}  
		\hline 
		   & \multicolumn{2}{c}{Trusted} & \multicolumn{2}{c}{Clickbait} &\multicolumn{2}{c}{Conspiracy} & \multicolumn{2}{c}{Propaganda} & \multicolumn{2}{c}{Disinfo}\\
		   
 	 Reaction 	&	$\Delta$ 	&	$\%\Delta$	&	 $\Delta$ 	&	$\%\Delta$	&	 $\Delta$ 	&	$\%\Delta$	&	 $\Delta$ 	&	$\%\Delta$	&	 $\Delta$ 	&	$\%\Delta$	\\
        \hline	 																				
Answer 	&	0.15	&	21.74	&	0.21	&	38.89	&	0.12	&	\textbf{19.05}	&	0.34	&	\textbf{\textit{58.62}}	&	0.11	&	\textbf{16.42}	\\
Elaboration 	&	0.09	&	\textit{18.00}	&	 --- 	&	---	&	 --- 	&	---	&	0.07	&	12.96	&	0.04	&	5.63	\\
Question 	&	0.12	&	25.53	&	0.13	&	\textbf{\textit{40.63}}	&	 --- 	&	---	&	0.16	&	34.78	&	0.05	&	9.80	\\
Other 	&	0.11	&	\textbf{28.95}	&	0.09	&	\textit{31.03}	&	 --- 	&	---	&	0.10	&	25.64	&	0.05	&	9.62	\\

\ignore{Answer 	&	0.15	&	21.74	&	0.21	&	38.89	&	0.12	&	19.05	&	0.34	&	58.62068966	&	0.11	&	16.42	\\
Elaboration 	&	0.09	&	18.00	&	 --- 	&	---	&	 --- 	&	---	&	0.07	&	12.96296296	&	0.04	&	5.63	\\
Question 	&	0.12	&	25.53	&	0.13	&	40.63	&	 --- 	&	---	&	0.16	&	34.7826087	&	0.05	&	9.80	\\
Other 	&	0.11	&	28.95	&	0.09	&	31.03	&	 --- 	&	---	&	0.10	&	25.64102564	&	0.05	&	9.62	\\
}

        \hline
 	\end{tabular}
 	 
 	\label{tab:gini_deltas}
 \end{table*}

Finally, we investigate reaction inequality to answer the question: does each user share an equal number of reactions, or are some user or users responsible for a disproportionate number of the reaction tweets for each of the most common reaction types (answer, elaboration, question, and ``other'') ? In Figure \ref{fig:lorenz} we present the Lorenz curves for bots and human users when we consider populations with reaction tweets for each combination of reaction type and class of source.

There are significant differences (MWU $p<0.01$) between the Lorenz curves for bot and human users for all combinations of reaction and source types except for elaboration reactions to clickbait news sources and elaboration, question, and ``other'' reactions to conspiracy sources. In these cases, human users are also unevenly responsible for reaction tweets, \ie  a subset of the human users are responsible for a disproportionate number of the human-reactions, and the disparity between users who react infrequently and those who post a substantial number of reactions is similar to those within the corresponding populations of bot users.

When users reacted to conspiracy sources, the volume of reaction tweets are similarly unequally distributed across users within the populations of bots and human users except for answer-reactions. Answer-reactions posted in response to conspiracy sources have a smaller prolific subset of bot users responsible for an unexpectedly large volume of the reaction tweets. Human users also respond unevenly with a subset of users who post a disproportionate amount of the reactions, but to a lesser extent than the population of bot users who posted reactions.  
We see similar patterns across all significant comparisons. That is, \textit{bot populations, if significantly different from the corresponding human user population, always have a higher level of disparity in reaction volumes than the corresponding human users}. 

Table~\ref{tab:gini_deltas} presents the increases in Gini coefficient from the human user to bot populations. For clarity, we present only the significant increases ($p < 0.05$) with dashes ($-$) in place of results without significance. Increases are presented in both absolute terms and relative to the Gini coefficient of the human user population. The most extreme difference is seen in answer-reaction to propaganda sources, with the bot population having a Gini coefficient 58.6\% (+0.34) larger than human users do. We find that the highest relative increases for the more deceptive news source classes (conspiracy, propaganda, and disinformation) occur when we compared answer-reactions. The highest relative increase for elaboration reactions occurs within elaboration-reactions to trusted news sources. The highest relative increase in inequality for reactions to trusted news source, however, occurs within the class of ``other'' reactions, \ie reactions that our annotation model did not predict to be one of the eight reaction types.
In contrast, we see the lowest significant relative differences between human and bot users in reactions to disinformation sources. We see that the Gini coefficients for bots are only 5.6\% higher than humans for elaboration-reactions, and approximately 10\% higher for both question-reactions and other-reactions.

\section{Conclusions and Future Work}
We have presented a novel analysis of bot and human-user reactions to sources of varying levels of credibility using fine-grained reaction labels. We identified several key differences in the prevalence of bots within reactions and populations of users who reacted, the variety of reactions each news source evokes, the speed with which different reactions occurred and the inequality of participation in the set of reactions. 
Future work will focus on further exploration of the differences in evolution of the response to deceptive sources, an expanded analysis that incorporates both frequent and infrequently reacting users, and comparisons across multiple platforms \eg Facebook and Reddit.

\section*{Acknowledgments}
Twitter data used for the analysis in this paper was collected using public Twitter API and analyzed over the period of 01/2016 -- 01/2017. Botometer data was collected by the University of Notre Dame using public APIs. The research was supported by the Laboratory Directed Research and Development Program at Pacific Northwest National Laboratory, a multiprogram national laboratory operated by Battelle for the U.S. Department of Energy. This research is also supported by the Defense Advanced Research Projects Agency (DARPA), contract W911NF-17-C-0094. The U.S. Government is authorized to reproduce and distribute reprints for Governmental purposes notwithstanding any copyright annotation thereon. The views and conclusions contained herein are those of the authors and should not be interpreted as necessarily representing the official policies or endorsements, either expressed or implied, of DARPA or the U.S. Government.

\bibliographystyle{IEEEtranS} 
\bibliography{deception_evolution}

\begin{thebibliography}{10}
\providecommand{\url}[1]{#1}
\csname url@samestyle\endcsname
\providecommand{\newblock}{\relax}
\providecommand{\bibinfo}[2]{#2}
\providecommand{\BIBentrySTDinterwordspacing}{\spaceskip=0pt\relax}
\providecommand{\BIBentryALTinterwordstretchfactor}{4}
\providecommand{\BIBentryALTinterwordspacing}{\spaceskip=\fontdimen2\font plus
\BIBentryALTinterwordstretchfactor\fontdimen3\font minus
  \fontdimen4\font\relax}
\providecommand{\BIBforeignlanguage}[2]{{%
\expandafter\ifx\csname l@#1\endcsname\relax
\typeout{** WARNING: IEEEtranS.bst: No hyphenation pattern has been}%
\typeout{** loaded for the language `#1'. Using the pattern for}%
\typeout{** the default language instead.}%
\else
\language=\csname l@#1\endcsname
\fi
#2}}
\providecommand{\BIBdecl}{\relax}
\BIBdecl

\bibitem{chu2010who}
Z.~Chu, S.~Gianvecchio, H.~Wang, and S.~Jajodia, ``Who is tweeting on twitter:
  Human, bot, or cyborg?'' in \emph{Proceedings of the 26th Annual Computer
  Security Applications Conference}, ser. ACSAC '10.\hskip 1em plus 0.5em minus
  0.4em\relax ACM, 2010, pp. 21--30.

\bibitem{davis2016botornot}
C.~A. Davis, O.~Varol, E.~Ferrara, A.~Flammini, and F.~Menczer, ``Botornot: A
  system to evaluate social bots,'' in \emph{Proceedings of the 25th
  International Conference Companion on World Wide Web}.\hskip 1em plus 0.5em
  minus 0.4em\relax International World Wide Web Conferences Steering
  Committee, 2016, pp. 273--274.

\bibitem{glenski2018identifying}
M.~Glenski, T.~Weninger, and S.~Volkova, ``Identifying and understanding user
  reactions to deceptive and trusted social news sources,'' in \emph{ACL},
  2018.

\bibitem{hargittai2008participation}
E.~Hargittai and G.~Walejko, ``The participation divide: content creation and
  sharing in the digital age,'' \emph{Information, Community and Society},
  vol.~11, no.~2, pp. 239--256, 2008.

\bibitem{howard2016bots}
P.~N. Howard and B.~Kollany, ``Bots,\# strongerin and\# brexit: Computational
  propaganda during the uk-eu referendum,'' \emph{Social Science Research
  Network}, 2016.

\bibitem{jin2013epidemiological}
F.~Jin, E.~Dougherty, P.~Saraf, Y.~Cao, and N.~Ramakrishnan, ``Epidemiological
  modeling of news and rumors on twitter,'' in \emph{Social Network Mining and
  Analysis}, 2013.

\bibitem{kakwani1973estimation}
N.~C. Kakwani and N.~Podder, ``On the estimation of lorenz curves from grouped
  observations,'' \emph{International Economic Review}, pp. 278--292, 1973.

\bibitem{kumar2014detecting}
K.~K. Kumar and G.~Geethakumari, ``Detecting misinformation in online social
  networks using cognitive psychology,'' \emph{Human-centric Computing and
  Information Sciences}, vol.~4, no.~1, p.~14, 2014.

\bibitem{kwon2017rumor}
S.~Kwon, M.~Cha, and K.~Jung, ``Rumor detection over varying time windows,''
  \emph{PloS one}, vol.~12, no.~1, p. e0168344, 2017.

\bibitem{kwon2013prominent}
S.~Kwon, M.~Cha, K.~Jung, W.~Chen, and Y.~Wang, ``Prominent features of rumor
  propagation in online social media,'' in \emph{Proceedings of ICDM}, 2013.

\bibitem{lazer2018science}
D.~M. Lazer, M.~A. Baum, Y.~Benkler, A.~J. Berinsky, K.~M. Greenhill,
  F.~Menczer, M.~J. Metzger, B.~Nyhan, G.~Pennycook, D.~Rothschild
  \emph{et~al.}, ``The science of fake news,'' \emph{Science}, vol. 359, no.
  6380, pp. 1094--1096, 2018.

\bibitem{metaxas2012social}
P.~T. Metaxas and E.~Mustafaraj, ``Social media and the elections,''
  \emph{Science}, vol. 338, no. 6106, pp. 472--473, 2012.

\bibitem{qazvinian2011rumor}
V.~Qazvinian, E.~Rosengren, D.~R. Radev, and Q.~Mei, ``Rumor has it:
  Identifying misinformation in microblogs,'' in \emph{Proceedings of EMNLP},
  2011.

\bibitem{rath2017retweet}
B.~Rath, W.~Gao, J.~Ma, and J.~Srivastava, ``From retweet to believability:
  Utilizing trust to identify rumor spreaders on twitter,'' in
  \emph{Proceedings of ASONAM}, 2017.

\bibitem{ratkiewicz2011detecting}
J.~Ratkiewicz, M.~Conover, M.~R. Meiss, B.~Gon{\c{c}}alves, A.~Flammini, and
  F.~Menczer, ``Detecting and tracking political abuse in social media.''
  \emph{ICWSM}, vol.~11, pp. 297--304, 2011.

\bibitem{shao2017spread}
C.~Shao, G.~L. Ciampaglia, O.~Varol, A.~Flammini, and F.~Menczer, ``The spread
  of fake news by social bots,'' \emph{arXiv preprint arXiv:1707.07592}, 2017.

\bibitem{starbird2017examining}
K.~Starbird, ``Examining the alternative media ecosystem through the production
  of alternative narratives of mass shooting events on twitter,'' in
  \emph{ICWSM}, 2017.

\bibitem{tambuscio2015fact}
M.~Tambuscio, G.~Ruffo, A.~Flammini, and F.~Menczer, ``Fact-checking effect on
  viral hoaxes: A model of misinformation spread in social networks,'' in
  \emph{WWW}, 2015.

\bibitem{vanmierlo2014rule}
T.~van Mierlo, ``The 1\% rule in four digital health social networks: An
  observational study,'' \emph{J Med Internet Res}, vol.~16, no.~2, p. e33, Feb
  2014.

\bibitem{varol2017online}
O.~Varol, E.~Ferrara, C.~A. Davis, F.~Menczer, and A.~Flammini, ``Online
  human-bot interactions: Detection, estimation, and characterization,'' in
  \emph{ICWSM}, 2017.

\bibitem{volkova2018misleading}
S.~Volkova and J.~Y. Jang, ``Misleading or falsification: Inferring deceptive
  strategies and types in online news and social media,'' in \emph{Companion of
  the The Web Conference 2018 on The Web Conference 2018}.\hskip 1em plus 0.5em
  minus 0.4em\relax International World Wide Web Conferences Steering
  Committee, 2018, pp. 575--583.

\bibitem{volkova2017separating}
S.~Volkova, K.~Shaffer, J.~Y. Jang, and N.~Hodas, ``Separating facts from
  fiction: Linguistic models to classify suspicious and trusted news posts on
  twitter,'' in \emph{ACL}, 2017.

\bibitem{Vosoughi1146}
\BIBentryALTinterwordspacing
S.~Vosoughi, D.~Roy, and S.~Aral, ``The spread of true and false news online,''
  \emph{Science}, vol. 359, no. 6380, pp. 1146--1151, 2018. [Online].
  Available: \url{http://science.sciencemag.org/content/359/6380/1146}
\BIBentrySTDinterwordspacing

\bibitem{wojcik2018pew}
S.~Wojcik, S.~Messing, A.~Smith, L.~Rainie, and P.~Hitlin, ``Bots in the
  twittersphere,''
  http://www.pewinternet.org/2018/04/09/bots-in-the-twittersphere/, 2018.

\bibitem{woolley2016automating}
S.~C. Woolley, ``Automating power: Social bot interference in global
  politics,'' \emph{First Monday}, vol.~21, no.~4, 2016.

\bibitem{wu2015false}
K.~Wu, S.~Yang, and K.~Q. Zhu, ``False rumors detection on sina weibo by
  propagation structures,'' in \emph{ICDE}.\hskip 1em plus 0.5em minus
  0.4em\relax IEEE, 2015.

\bibitem{wu2016mining}
L.~Wu, F.~Morstatter, X.~Hu, and H.~Liu, ``Mining misinformation in social
  media,'' \emph{Big Data in Complex and Social Networks}, 2016.

\bibitem{zhang2017characterizing}
A.~Zhang, B.~Culbertson, and P.~Paritosh, ``Characterizing online discussion
  using coarse discourse sequences,'' in \emph{ICWSM}, 2017.

\bibitem{zhao2015enquiring}
Z.~Zhao, P.~Resnick, and Q.~Mei, ``Enquiring minds: Early detection of rumors
  in social media from enquiry posts,'' in \emph{Proceedings of WWW}, 2015.

\end{thebibliography}

\end{document}